\documentclass{ws-procs9x6}

\def\be{\begin{equation}}
\def\ee{\end{equation}}
\def\bea{\begin{eqnarray}}
\def\eea{\end{eqnarray}}

\begin{document}

\title{S-Branes, Negative Tension Branes and Cosmology}

\author{F.~Quevedo$^1$, G.~Tasinato$^2$ and I.~Zavala C$^1$.}

\address{$^1$Centre for Mathematical Sciences, DAMTP, University of 
          Cambridge,\\
        Cambridge CB3 0WA, U.~K. \\
        E-mail: F.Quevedo@damtp.cam.ac.uk, 
                eiz20@damtp.cam.ac.uk \\
        $^2$Physikalisches Institut der Universit\"at Bonn, \\
    Nussallee 12, 53115 Bonn, Germany.\\ 
        E-mail: tasinato@th.physik.uni-bonn.de}

\maketitle

\abstracts{A general class of solutions of string background 
 equations 
is studied and
its physical interpretations are  presented. These solutions correspond to
generalizations of the standard black p-brane solutions to
 surfaces with
curvature $k=-1,0$. The relation with the recently introduced S-branes
is provided. The mass, charge, entropy and Hawking temperature are
computed, illustrating the interpretation in terms of negative tension
branes. Their cosmological interpretation is discussed as well as their
potential instability under small perturbations. 
(Work done in collaborations with C.~Grojean, and with C.~P.~Burgess
 and S.~J.~Rey). 
%
}

\section{Motivations and Conclusions}

One of the most interesting ideas coming out of the brane world
scenario is what is known as `mirage cosmology'
\cite{rev}. A 3-brane moving in a 
(static) 5d anti de Sitter black hole induces, for an observer on the
brane, the illusion that his 4d universe is expanding. In this case,
the treatment is completely equivalent to the standard cosmological
description of the brane as a fixed object embedded in a
time-dependent  background. The equivalence
is due to the existence of Birkhoff's theorem in 5d, which essentially states
that the most general spherically symmetric metric is static.

This view is very  interesting, and we may wonder if it also applies to
 backgrounds more stringy than just 5d anti de Sitter space. The generalization
can go in several directions. First, we should start with an action
including the typical bulk fields of string theory: the graviton,
dilaton and antisymmetric RR or NS-NS fields in d-dimensions. Second,
we have to include solutions that go beyond the spherical symmetric
 black hole  solution, which  would include only FRW
cosmologies with curvature $k=1$. We then need to look for black
hole-like solutions with curvatures $k=-1,0$. Third, we have to see if
Birkhoff's theorem holds in these more general settings.

Another, {\it a posteriori},  motivation for the analysis of this kind of
 solutions, is that they present alternative
cosmologies to the standard big-bang, in which, extrapolating the scale
factor to the past, it does not meet a singularity but a regular
 horizon for the metric. 
The possible realisation of these geometries would have very
important implications for early universe cosmology
\cite{gqtz,bqrtz,costa}.

Moreover, there has been recent interest in the study of space-like or 
S-branes in string theory~\cite{sbranes}, as time-dependent solutions
of supergravity equations.
 Their interpretation 
offers  important challenges, like the understanding of what kind of
physical objects
these space-times correspond to. In particular, in their original 
definition, there was no reference to the concept of conserved charges, and
the study of their  stability under perturbations was not considered.

We will start by presenting  the analysis of solutions to the
Einstein-dilaton-antisymmetric tensor equations with any curvature
$k=1,-1,0$. The $k=1$ case corresponds to standard black
$p$-branes. For $k=-1,0$ we have a geometry with past and
future cosmological regions, separated, by regular horizons, from
static  regions. The analysis of these static regions  reveal the
presence of  time-like singularities: we will provide a physical
interpretation for the singularities 
 in terms of a pair of negative tension branes, with opposite
charge.  The solutions describe also S-brane--like objects, sited on
the  horizons. 
The similarity with the Schwarzschild
black hole allows us to compute the Hawking temperature and
entropy
density. The stability of the whole solution and the horizons is
discussed: the past horizons appear to be unstable, due to 
infinitely blue-shifted energy flux coming from the past cosmological region.

We finally note that the mirage cosmology cannot have the level of
generality as in the simple 5d case, since Birkhoff's theorem does not
hold in the presence of a dilaton. In particular, this implies that
the  cosmology of
a brane world will have two different sources, the time dependence of the 
bulk background and the motion of the brane through this  background.

\begin{figure}[th]
\centerline{\epsfxsize=1.6in\epsfbox{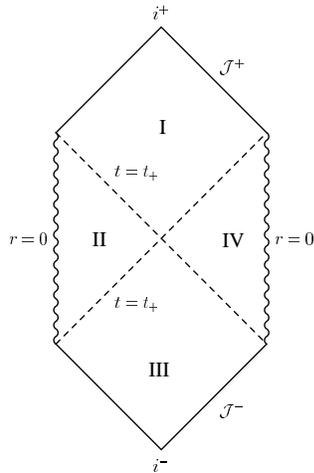}}   
\caption{Penrose diagram for our time-dependent solutions. Notice that it correspond to the conformal diagram for a Schwarzschild black hole, but {\it rotated} by $90^o$ degrees. \label{sbrane}}
\end{figure}

\section{General solutions}\label{Sec3}

We now turn to the description of  various
space-time and brane's world-volume dimensions, for a system 
involving metric, dilaton, and $(q+1)$-form tensor fields --- a
system encompassing bosonic fields of diverse supergravity or
superstring theories and their compactifications. The class
of solutions we present, has been obtained in~\cite{gqtz,bqrtz}.

\subsection{Dilaton-Antisymmetric tensor-Einstein solutions}\label{sec3.1}
We are interested in the following
Einstein-frame action, in $d=(n+q+2)$-dimensional space-time:
\begin{equation}\label{generalaction}
S= \frac{1}{2}\int_{\mathcal{M}_d} d^{d}x \sqrt{g}
\left[ R -  (\nabla\phi)^2 -
\frac{1}{(q+2)!} e^{-\sigma \phi} F_{q+2}^2 \right],
\end{equation}
where $g_{\mu \nu}, \phi, F$ denotes metric, dilaton field, and
$(q+2)$-form tensor field strength, respectively. 
 Eq.~(\ref{generalaction}) includes supergravity, and so also
low-energy string theory, for specific choices of $d$, $\sigma$
and $q$.

We are   interested in classical solutions whose space-time geometry
takes the form of an asymmetrically warped product between
$q$-dimensional flat space-time and $n$-dimensional
maximally-symmetric space, characterized by its  curvature $k
= 0, \pm 1$. That is,  solutions that depend only on one
warping variable --- either $t$ or $r$. The Ansatz is suitable to  
describe a flat $q$-brane propagating in
$d=(n+q+2)$-dimensional ambient space-time, where $n$-dimensional
transverse hyper-surface is a space of maximal symmetry.
The solutions satisfying these requirements are given by
\bea
ds^2  & = &  h_-^A \left( -  h_+h_-^{1-(n-1) b}dt^2 +  h_+^{-1}h_-^{-1+b}
dr^2 +r^2h_-^{b} dx^2_{n,k} \right) + h_-^B dy^2_{q}\,,
\qquad \label{metric}\\
\phi &=& \frac{(n-1) \sigma b}{\Sigma^2} \ln h_- \,,\label{eq:dil} \\
F_{try_1 \dots y_q} &=& Q \epsilon_{try_1 \dots y_q} r^{-n},
\qquad \epsilon_{try_1 \dots y_q}=\pm 1 \,.\label{eq:F}
\end{eqnarray}
The notations are as follows. The metric of the $n$-dimensional
maximally symmetric space, whose Ricci scalar is equal to $n(n-1)k$
for $k=0,\pm 1$, is denoted with $d x^2_{n, k}$. The harmonic
functions $h_\pm(r)$ depend on two first-integral constants,
$r_\pm$, and are given by:
\begin{equation}\label{h+}
h_+(r) = \pm \left(1-\left(\frac{r_+}{r}\right)^{n-1} \right),
\qquad h_-(r) = \left|k-\left(\frac{r_{-}}{r}\right)^{n-1}\right|,
\end{equation}
where $\pm$ refers to the sign of $h_-$.
The constant $Q$ is given by
\begin{equation}\label{Qdef}
Q = \left(\frac{n(n-1)^2
(r_+r_-)^{n-1}}{n \Sigma^2 +
4(n-1)} \right)^{1/2},
\end{equation}
where $\Sigma$ and $b$ are:
\begin{eqnarray*}
\Sigma^2 = \sigma^2 + {4}
\frac{q(n-1)^2}{n(n+q)}\,, \qquad b = \frac{2 n
\Sigma^2}{(n-1)( n \Sigma^2+ 4(n-1))}\,.
\end{eqnarray*}
Likewise, the exponents $A, B$ in eq.~(\ref{metric}) are given by:
\begin{eqnarray*}
A = - \frac{4  q (n-1)^2 b }{ n (n+q) \Sigma^2}\,,
\qquad \mbox{and} \qquad B =  -\frac{n}{q} A\,.
\end{eqnarray*}


The two integration constants, $r_\pm$, are related intimately
to two conserved charges associated with the solution. One of
these is the $q$-form electric charge $Q$ --- see eq.~(\ref{eq:F})
--- acting as the source of the $(q+2)$-form tensor field
strength, while  the other will correspond to the
 mass of the object in a form that we
will discuss later.


\smallskip

Let us now describe the  geometrical characteristics
of these solutions. First of all, it is not hard to find that, for any $k$,
$r_+$ is always a horizon. The other properties depend
on the value of the constant curvature $k$:
%
%

\smallskip

\noindent $\bullet$ \textbf{\textit{k=1:}}
By computing the scalar curvature, it is possible to realize that $r=r_-$
is  a scalar singularity for any positive value of $b$.\footnote{For
$b=0$, $r=r_-$
is actually another horizon as in the Reissner--Nordstr\"om solution.}
Then in order to avoid the presence of naked singularities, one
requires that $r_-<r_+$ and, moreover, we should also require 
 $(r_-r_+)^{n-1}\ge 0$ in order to have a real valued charge. 
The background is asymptotically flat and corresponds (for $d=10$),
to the black $q$-brane solutions constructed by Horowitz and  
Strominger~\cite{HS}.  
%

\noindent $\bullet$ \textbf{\textit{k=0 and k=--1:}}
One finds, from the expressions for $h_-$ above, that $r=r_-$ is  a
regular point for the metric. The coordinate
$r$ becomes time-like in the region $r>r_{+}$
 while remains space-like for $r<r_+$, exactly the opposite of
Schwarzschild black  hole.
The $q$-dimensional singularity at $r=0$ is then time-like. The Penrose
diagram  is presented in  Fig. (\ref{sbrane}): each
point  correspond to an $n$ dimensional hyperboloid (for $k=-1$) or an $n$
 dimensional flat space (for $k=0$), cross a $q$
dimensional  flat surface\footnote{It is interesting to observe that
the  solution for $k=-1$ can be obtained by
 an analytical continuation of the $k=1$ case~\cite{bqrtz}.}.
 These solutions do not correspond to black
objects, but then, what are they? We will discuss this point in what
follows.


\section{Properties and Interpretation}


 It is possible to interpret our solutions, for $k=0,-1$ and $q=0$,
 as particular examples 
 of  space-like branes (S-branes), as described recently in~\cite{sbranes}. An S-brane is 
a brane-like object for which the world-volume is space-like. S-branes
 can be found as time-dependent
solutions of supergravity equations. 

In the original S-brane solutions, the  static
regions were ignored: it was then not possible to define conserved
quantities such as charge or mass for those objects. We will see, in
what follows, how the inclusion of those regions allows us to
define various properties for these solutions and, moreover, give us an
interesting interpretation for this geometry in terms of negative tension objects. 

It is also interesting to write the solutions in terms of  the
 conformal time in which the nature of a  time-like sort of kink,
 associated to the S-branes,
becomes evident and, moreover, a cosmological bouncing evolution is
also clear. 
 Let us consider for concreteness the case $k=-1$ and $q=0$, in the absence of scalar and antisymmetric form.
 In this case, the metric in the original
coordinates is
\begin{eqnarray}
d s^{2} = - \frac{1 }{h_{+}} d t^{2}+ h_{+}\, d r^{2} +
t^{2} d x_{n,-1}^2\,,
\label{simplif}
\end{eqnarray}
where $h_{+}=1-({r_{+}}/{t})^{n-1} $. We now rewrite this metric
in terms of the conformal time $\eta$  as
\begin{eqnarray}
d s^{2} = C^2(\eta)\left[- d\eta^2 + d x_{n,-1}^2\right]
+ D^2(\eta)dr^2\,,
\end{eqnarray}
where the conformal time is defined by
\begin{eqnarray}
C(\eta)= t(\eta) =
r_+ \cosh ^{2/(n-1)}\left[\frac{(n-1)}{2} \eta \right] \ge r_+\,,
\end{eqnarray}
and so $\eta$ ranges over $ -\infty < \eta < \infty $. The
scale factor for the $r$-direction becomes
\begin{eqnarray}
D(\eta) = \tanh \left[\frac{(n-1)}{2}\eta \right]
\end{eqnarray}
and has the same functional dependence for any value of $n$.
These expressions exhibit a bouncing structure of the
$(n+1)$-dimensional space and a (time-like) kink structure for
the radial dimension.


\subsection{Conserved quantities}

We now  identify two conserved quantities as Noether
charges, carried by the  branes at the singularities, whose metric, dilaton field,
and $(q+2)$-form field strength are given  in
eqs.~(\ref{metric})--(\ref{eq:F}). 

\paragraph{Electric charge.}


From the field equation eq.~(\ref{eq:F}) of the $(q+2)$-form
tensor field strength, a conserved charge density can be defined
through $d^*F_{q+2} = {}^*J$. This leads to the following
expression for the electric charge:
\begin{equation}\label{Qint}
Q=\int_{\Sigma} d\Sigma_{\mu i\dots}\nabla_{\nu}
\left( e^{-\sigma\phi} F^{\mu \nu i \dots}\right) =
\int_{\partial\Sigma} d\Sigma_{\mu\nu i\dots} e^{-\sigma\phi}
F^{\mu\nu i \dots}\,,
\end{equation}
where $\Sigma$ refers to any $(n+1)$-dimensional space-like
hyper-surface transverse to the $q$-brane. The main advantage of the above
expression for the electric charge lies in the observation that the
integrand vanishes almost everywhere by virtue of the field
equation for $F_{q+2}$. It does not vanish literally everywhere,
 because the integrand behaves like a delta function at
each of the two time-like singularities. Conservation of $Q$ is
also clear in this formulation, as the second equality of
eq.~(\ref{Qint}) shows that $Q$ does not depend on $\Sigma$ as
long as the boundary conditions on $\partial \Sigma$ are not
changed.

We are led in this way to identify the conserved quantities, $\pm
Q$, with electric charges carried by each of the two $q$-branes
located at the time-like singularities (which, unlike the horizons,
 {\it are not} S-branes). Which brane carries which
sign of the electric charge may be determined as follows. As
eq.~(\ref{eq:F}) defines the constant $Q$ relative to a coordinate
patch labeled by $r$ and $t$, the key observation
is that the coordinate $t$ can increase into the
future only for one of the two regions, II or IV. Then, the
charge $+Q$ applies to the brane whose static region $t$
increases into the future, and $-Q$ applies to the brane whose
$t$ increases into the past.

\paragraph{Gravitational mass.}

We can also associate a mass to these objects by adopting  the  
\emph{Komar integral}
formalism, which cleanly associates a conserved
quantity with a Killing vector field, $\xi^\mu$, by defining a flux
integral (see however~\cite{marija}):
\begin{eqnarray} \label{Mdef}
K[\xi] := \frac{c}{16\pi G} \oint_{\partial
\Sigma} {dS_{\mu\nu}D^\mu \xi^\nu}\,.
\end{eqnarray}
Here, $c$ denotes a normalization constant, and $\Sigma$ is again
an $(n+1)$-dimensional space-like hyper-surface transverse to the
$q$-brane, and $\partial \Sigma$ refers to the boundary of
$\Sigma$. We have also recovered units by reintroducing Newton's
constant, $8\pi G$.  The Komar charge $K$ is manifestly conserved, since it is
invariant under arbitrary deformations of the space-like
hyper-surface $\Sigma$ for a fixed value of the fields on the
boundary $\partial \Sigma$.

To evaluate the tension $\mathcal{T} = K[\partial_t]$ in the
patch for which $\partial_t$ is future-directed, we will choose
for the hyper-surface $\Sigma$ a constant-$t$ spatial slice and
for the boundary $\partial\Sigma$ a $r =r_0$ (viz. a constant
radius) slice in the regions II and IV, respectively. It turns
out that, if $Q\ne 0$, the expression for the tension depends on
the value $r$ at which the boundary $\partial \Sigma$ is
defined\footnote{The same  happens  for the radius-dependent mass in
Reissner-Nordstr\"om black-holes.}. 
Explicitly, we find the tension is given by:
\begin{eqnarray}
\frac{\mathcal{T} (r)}{V} =
 -\frac{(n-1)}{8\pi G}\left[ r_-^{n-1} -
k r_+^{n-1} + \frac{2 Q^2 }{(n+q)(n-1)} \left( \frac{1}{r^{n-1}} -
\frac{1}{r_+^{n-1}} \right) \right]\,
\label{tensione}
\end{eqnarray}
Negative tension, $\mathcal{T}<0$, for both branes is in accord
with the form of the Penrose diagram of Fig.(~\ref{sbrane}), which,
in the static regions, II and IV, is similar to the Penrose
diagram for a negative-mass Schwarzschild
black-hole, or to the over charged region of
the Reissner-Nordstr\"om black-hole.
Negative-valued gravitational mass or tension is also borne out
by the behavior of time-like geodesics  in these
regions~\cite{bqrtz}.

\subsection{Thermodynamics}

Given the explicit time dependence of the space-time in the
time-dependent regions I and III, one would expect that particle
production takes place in these regions.   One can  show that a
Hawking temperature can be associated to the horizons
 with the static regions II and IV of the space-time.

\paragraph{Hawking temperature.}
To estimate  the temperature we proceed in the usual way, 
 performing a Euclidean continuation of
the metric in the static  region by sending $t\rightarrow i\tau$, and
then demanding not to have a conical singularity on the horizon in this
Euclidean space-time. This condition requires the Euclidean time
coordinate to be periodic $\tau \sim \tau + 2 \pi/\kappa$, and so
implicitly defines a temperature as:
\begin{eqnarray}
T = \frac{\kappa}{2 \pi} = \frac{n-1}{4 \pi r_+}
\left|k -\left(\frac{r_-}{r_+} \right)^{n-1} \right|^{1 - nb/2}.
\end{eqnarray}
This reduces to previously obtained expressions for the special
cases where these metrics agree with those considered elsewhere.
In particular, it vanishes for extremal black-branes, for which $k
= 1$ and $r_- = r_+$.

\paragraph{Entropy.}
The possibility to define the temperature of a space-time
involving horizons immediately suggests that it may also be
possible to associate it an entropy. In fact by following a
standard procedure computing the Euclidean action,  one arrives to
a simple expression for the entropy\cite{bqrtz}
\begin{equation}
s=\frac{S}{V} = \frac{r_+^{n}}{4 G}\left| k-
\left(\frac{r_-}{r_+}\right)^{n-1}
\right|^{nb/2} = \frac{1}{4G} \sqrt{g_{nn}}\left|_{r_+} \right.,
\end{equation}
It is remarkable that, for any $k$, the expression for the entropy
does not depend on the coordinate $r: g_{nn}$ corresponds to
the determinant of the induced metric on the $n$ spatial
dimensions,  calculated at the horizon $r_{+}$. 

In the case
$k=1$, we obtain the well-known relation
\begin{equation}
S=\frac{\mathcal{A}}{4G}\,,
\end{equation}
where $\mathcal{A}$ is the area of the black-hole horizon. 
For $k=-1$ and 0,  the area of the horizon is infinite, but we can
still consider the entropy per unit volume and the same formula holds.

\subsection{Stability}
The stability of the whole solution is  difficult to
establish. Considering       the simple case of a massive scalar field
embedded in the space-time and satisfying the
Klein Gordon equation, we  have checked that it does not present
fast  growing   modes: this indicates that the time-dependent regions
are  stable under perturbations.
However, the energy flux of
the modes of the scalar blows up at the past Cauchy horizon,
suggesting  a  possible instability  similar to the one of the  
Reissner-Nordstr\"om black hole.

\section{Realistic Cosmology?}
These solutions, with a contracting  phase followed by 
an expanding one, without meeting a singularity,  constitute interesting
cosmological backgrounds.
However,
there are obstacles to promote this kind of solutions to a realistic
cosmological setting.
First, the instability of the Cauchy horizon may re-introduce the initial
singularity. Furthermore, in 4D these solutions are not isotropic.
To have 4D isotropy we have to consider higher dimensional solutions, and
 some of the spatial  dimensions need to be compactified. This
procedure  generally implies the presence of regions with closed
time-like  curves, and/or some instability under small perturbations~\cite{hp}.

On the other hand, we may construct  brane world models in this background.
Introducing a moving brane in the asymptotically flat regions, 
we have  an induced expansion on the brane coming both from
the cosmological expansion and the motion of the brane. In particular, the 
induced Friedmann's equation on the brane will present the usual brane-world 
form
$H^2\sim \rho^2+\cdots $. 
The  procedure of adding a constant
$\Lambda$
 to
the energy density,  and then expand the square to get a linear term in
$\rho$ plus small corrections, does not work here.  This because, in
the time  dependent region,
 the $\Lambda^2$ term can not be compensated by any bulk contribution.
 The resulting model describes then an accelerating universe with
positive  cosmological constant.
As a final comment, notice that
 Birkhoff's
theorem does not hold~\cite{gqtz,bgqtz},
 due to the presence of the scalar field: the most general maximally
symmetric metric will depend (at least) on two 
coordinates. Therefore,
to study the most general cosmological configuration, 
solutions depending on $r$ and $t$ should be found. We can see there are 
many interesting challenges in this field.

\vskip 0.2 cm
We thank C.P. Burgess, C. Grojean and S.-J. Rey for very enjoyable 
collaborations. F.~Q.~is partially supported by PPARC. G.~T.~is
supported by the European TMR Networks HPRN-CT-2000-00131, 
HPRN-CT-2000-00148 and  HPRN-CT-2000-00152. I.~Z.~C.~ is supported by
CONACyT, Mexico.

\end{document}